\documentclass[prb,twocolumn,notitlepage,longbibliography]{revtex4-2}
\usepackage{amsmath}
\usepackage{amssymb}
\usepackage{bbold} 
\usepackage{natbib}
\usepackage{lipsum}
\usepackage[unicode=true,colorlinks=true,citecolor=blue,urlcolor=blue]{hyperref}

\usepackage{bm}
\usepackage{epsfig} 
\usepackage{tikz} 
\usepackage[normalem]{ulem}

\renewcommand {\Im}{\mathop\mathrm{Im}\nolimits}
\renewcommand {\Re}{\mathop\mathrm{Re}\nolimits}

\renewcommand {\phi}{{\varphi}}
\newcommand {\rmi}{{\rm i}}

\newcommand {\e}{{\rm e}}

\begin{document}
\title{Resonant enhancement of second harmonic generation in 2D nonlinear crystal integrated with meta-waveguide: analytical vs numerical approaches}

\author{Egor S. Vyatkin}
\affiliation{Ioffe Institute, St. Petersburg 194021, Russia}

\author{Sergey A. Tarasenko}
\affiliation{Ioffe Institute, St. Petersburg 194021, Russia}

\email{}

\begin{abstract}
We present an analytical theory of second harmonic generation (SHG) in hybrid structures combining a nonlinear 2D crystal with a dielectric metasurface waveguide. The theory describes the excitation spectrum and enhancement of SHG at both leaky mode and quasi-bound state in the continuum (quasi-BIC) resonances in terms of the material parameters. For low-loss systems, the SHG efficiency at leaky resonances is determined by their radiative broadening, governed by the relevant Fourier harmonics of the metasurface polarizability, whereas the SHG enhancement at quasi-BIC resonances is ultimately limited by inhomogeneous broadening and absorption in the system. We also describe the emergence and polarization properties of second harmonic diffracted beams. These beams appear even if both the 2D crystal and the meta-waveguide are centrosymmetric owing to the nonlocal mechanism of SHG. The developed framework provides a systematic theoretical basis for optimizing the resonant nonlinear frequency conversion in hybrid 2D-material–metasurface platforms and identifies the fundamental limitations of the SHG efficiency.

\end{abstract}
\date{\today}

\maketitle
%%%%%%%%%%%%%%%%%%%%%%%%%%%%%%%%%%%%

\section{Introduction}\label{sec:intro}

Dielectric metasurfaces fabricated from high-refractive-index and low-loss materials support sharp optical resonances associated with the excitation of localized photonic modes~\cite{Kuznetsov2016,Tikhodeev2002,Huang2023}. At the resonances, the near electromagnetic field at the fundamental frequency $\omega$ is enhanced, leading to a dramatic increase in nonlinear optical phenomena such as second harmonic generation (SHG)~\cite{Grinblat2021,Wang2025}. Research in this field is at the core of modern nanophotonics and currently being actively pursued, both experimentally and theoretically, for metasurfaces made of nonlinear materials~\cite{Liu2016,Ling2025} and for hybrid structures combining dielectric metasurfaces with nonlinear 2D crystals~\cite{Bernhardt2020,Ning2023}.
Of particular interest for efficient frequency conversion is the realization of double resonance, where the electromagnetic field at both the fundamental and second harmonic frequencies satisfies the resonant conditions~\cite{Celebrano2015}. Another approach to achieving giant SHG enhancement involves the use of extremely narrow resonances~\cite{Koshelev2020}. Such resonances originate from optically inactive bound states in the continuum (BICs), which, due to symmetry breaking, get coupled to the incident field and transform into quasi-BICs with exceptionally high but finite Q-factors~\cite{Koshelev2018,Xu2019,Dyakov2021,Kazarinov1976,Koshelev2021}.

\begin{figure}[t]
    \centering
    \includegraphics[width=0.6\linewidth]{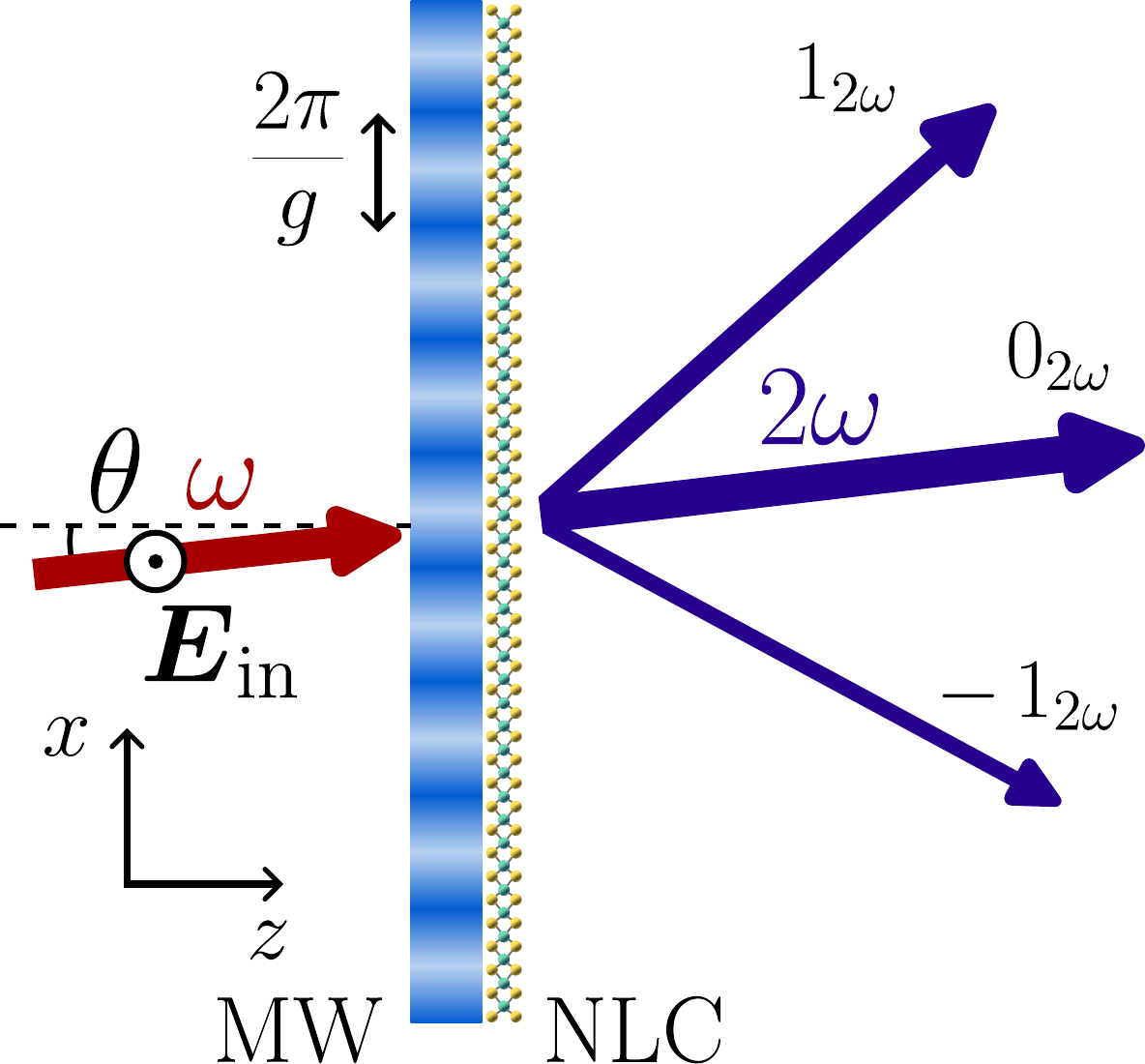}
    \caption{Second harmonic generation in 2D nonlinear crystal (NLC) integrated with dielectric metasurface waveguide (MW). Resonant excitation of bright (leaky) and dark (quasi-BIC) guided modes by the incident field at the fundamental frequency $\omega$ leads to the enhancement of the $2\omega$ emission. The $2\omega$ emission in the forward direction may include not only the collinear beam ''$0_{2\omega}$'' but also the diffracted beams ''$\pm 1_{2\omega}$''.}
    \label{fig:scheme}
\end{figure} 

Theoretical studies of SHG in metasurfaces are typically focused on modeling the electric field distribution and determining the spectral positions and widths of the Fano resonances from numerical calculations. Here, we present a microscopic analytical theory of resonant SHG, which enables the description of SHG in terms of material parameters and allows direct comparison with experiments and full calculations. We consider a hybrid structure where a 2D nonlinear crystal (NLC) is attached to a dielectric metasurface waveguide (MW), as shown in Fig.~\ref{fig:scheme}. This design combines the advantages of strong nonlinear response of 2D NLCs, such as graphene~\cite{Dean2009,Hendry2010,Glazov2011,Golub2014,Stepanov2020}, transition-metal dichalcogenides~\cite{Li2013,Mennel2018,Zhou2020}, and twisted van der Waals structures~\cite{Hsu2014,Yang2020,Yao2021,Paradisanos2022}, with the independent control of photonic modes in the MW~\cite{Meng2021,WangJ2025,Lou2021,Zang2021,Salakhova2023,Vyatkin2025}. We consider two sources of SHG: the standard mechanism related to the local second-order nonlinear susceptibility of the noncentrosymmetric NLC, and the nonlocal mechanism originating from the spatial inhomogeneity of the electromagnetic field in the 2D plane~\cite{Wang2009,Durnev2022,Gunyaga2025}. The latter mechanism is important here due to the strong spatial modulation of the near field in the NLC plane and predicts the emergence of second harmonic diffracted beams even when both the MW and NLC are centrosymmetric. 
We study the resonant enhancement of SHG at both leaky and quasi-BIC modes, the latter can be excited by a small deviation of the light incidence angle from normal or in an MW with broken inversion symmetry. The analytical approach allows us to describe SHG in terms of the material parameters, also in the presence of non-radiative resonance broadening, and to determine the fundamental limits of the SHG efficiency.
 
\section{Model and Theory}\label{sec:model}

Consider a 2D nonlinear crystal attached to a metasurface waveguide, which is a thin dielectric slab with a lateral modulation of the permittivity, Fig.~\ref{fig:scheme}. The permittivity is modulated along the $x$ direction, and the slab is characterized by the 2D polarizability $\alpha(x)$ with the Fourier harmonics
\begin{equation}\label{eq:alpha}
\alpha(x)=\sum_{n=-\infty}^{\infty}\alpha_n \e^{\rmi n g x}\,,
\end{equation}
where $2\pi / g$ is the modulation period. It is assumed that the modulation period is smaller than the wavelength of the incident light, i.e., $g > \omega / c$, so that the dielectric grating does not cause the diffraction of the incident light at the fundamental frequency $\omega$ (the so-called meta-waveguide regime). In the analytical calculations below, we consider a meta-waveguide with a low-contrast grating, $|\alpha_n| \ll \operatorname{Re}\alpha_0$ for $n \neq 0$~\cite{Bulgakov2018}, weak absorption, $\operatorname{Im} \alpha_0 \ll \operatorname{Re} \alpha_0$, and neglect the small absorption modulation, $\alpha_n^*=\alpha_{-n}$ for $n \neq 0$.
MWs with high $\operatorname{Re}\alpha_0$ and low-contrast gratings exhibit sharp optical resonances that enable high SHG efficiency.

A thin dielectric MW supports TE guided modes that can be excited by incident light with the appropriate polarization. 
% due to the dielectric grating. 
Accordingly, we consider the incident field with the wave vector $\bm q = (q_x,q_z)=(\omega/c)(\sin \theta, \cos \theta)$ and the polarization $\bm E_{\rm in} \parallel y$, which efficiently couples to the guided modes. The resonant excitation of these guided modes results in the formation of a strong near field $E_{y}(x)$, whose amplitude is much larger than that of the incident field $E_{\rm in}$. The local field, oscillating at the fundamental frequency, is converted into the field at the doubled frequency in the nonlinear crystal. 

For the geometry under study, the polarization at the doubled frequency $\bm P^{(2)}$ induced in the NLC is given by
\begin{align}\label{eq:P2}
P^{(2)}_x = \chi_{xyy} E_{y}^2 + \chi' E_{y}\frac{dE_{y}}{dx}\,, \quad
P^{(2)}_y = \chi_{yyy} E_{y}^2\,,
\end{align}
where $\chi_{xyy}$ and $\chi_{yyy}$ are components of the 2D second-order nonlinear susceptibility of the NLC, and $\chi'$ is a parameter describing the SHG due to the spatial inhomogeneity of the field~\cite{Gunyaga2025}. While the tensor $\chi$ requires the absence of inversion symmetry in the NLC, the parameter $\chi'$ is nonzero in any 2D materials, including centrosymmetric ones. 
The latter contribution is included because the near field is inherently non-uniform, being modulated with the grating period.

The modulation of the local field $E_y$ along $x$ leads to the modulation of the polarization $\bm P^{(2)}$ with the same lateral periodicity, Eq.~\eqref{eq:P2}.
Therefore, the far field at $2\omega$ in the forward direction may include not only the beam "$0_{2\omega}$" with the wave vector $2 \bm q$ but also the beams "$\pm 1_{2\omega}$" with the wave vectors $(2 q_x \pm g, \sqrt{(2\omega/c)^2-(2q_x \pm g)^2})$ provided $\omega/c > |q_x \pm g/2|$, as illustrated in Fig.~\ref{fig:scheme}.
The corresponding  angles of the emission at small incidence angle $\theta$ are given by
\begin{align}
    \tan \theta_{\pm1}=&\frac{g}{\sqrt{(2\omega/c)^2-g^2}}\\ \nonumber
    &\pm\frac{\theta\omega/c}{\sqrt{(2\omega/c)^2-g^2}}\bigg[2+\frac{g \omega/c}{(2\omega/c)^2-g^2}\bigg] \,.
\end{align}
The beams "$\pm 1_{2\omega}$" correspond to the first-order diffraction of the second-harmonic radiation or, equivalently, the half-order diffraction of the incident radiation.

To calculate the polarization $\bm P^{(2)}$ and the emitted field at $2\omega$ we first determine the local field at the fundamental frequency in the framework of the linear response theory. The spatial distribution of the field $E_y(x,z)$ is found from the wave equation 
\begin{equation}\label{eq:wave}
    \frac{\partial^2 E_y}{\partial x^2}+\frac{\partial^2 E_y}{\partial z^2}+\omega^2 E_y=-4\pi \omega^2\alpha(x)\delta(z)E_y\,,
\end{equation}
where we set $c=1$.
Solution of Eq.~\eqref{eq:wave} has the form
\begin{equation}\label{eq:Ey}
    E_y=\e^{\rmi q_x x}\bigg[E_{\rm in}\e^{\rmi q_z z}+r E_{\rm in}\e^{\rmi q_z |z|}+\sum_{n\neq0}E_n\e^{-\varkappa_{n}|z|+\rmi gnx}\bigg]\,,
\end{equation}
where $E_{\rm in}$ is the incident field amplitude, $r$ is the amplitude reflection coefficient, $E_n$ are the amplitudes of the Fourier   harmonics of the field at $z=0$, and $\varkappa_n (\omega, \theta)=\sqrt{(ng+\omega \sin \theta)^2-\omega^2}$ are the inverse decay lengths of the near field. The harmonics with $n\neq0$ correspond to the evanescent waves bound to the meta-waveguide. The transmission coefficient is given by $t=1+r$. The amplitudes $E_0=(1+r)E_{\rm in}$ and $E_n$ ($n\neq0$) are determined from the set of linear equations
\begin{align}\label{eq:En}
    E_0&=\frac{2\pi \rmi \omega}{\cos\theta}\sum_{m}\alpha_m E_{-m}+E_{\rm in}\,,\\ \nonumber
    E_{n}&=\frac{2\pi\omega^2}{\varkappa_{n}}\sum_{m}\alpha_m E_{n-m}\,.
\end{align}

Equations~\eqref{eq:Ey} and~\eqref{eq:En} fully determine the spatial distribution of the field $E_y(x,z)$ including the near field structure. Spectrally far from the MW resonances, $|E_n|\ll|E_0|$, and the field distribution resembles that of a uniform slab. 
% The situation changes dramatically 
At the MW resonances, certain Fourier harmonics of the near field $E_n$ become much larger than $E_0$. 
Such a resonance enhancement occurs at $\omega\approx\Omega(\omega\sin\theta+ gn)$, where $\Omega(k)$ is the guided mode dispersion. For a uniform slab with the polarizability $\alpha_0$, the dispersion reads
\begin{equation}
    \Omega(k)=\frac{1}{2\sqrt{2}\pi \alpha'_0}\sqrt{\sqrt{1+(4\pi \alpha'_0 k)^2}-1}\;,
\end{equation}
where $\alpha'_0 = \operatorname{Re} \alpha_0$. Note that the small imaginary part $\alpha''_0 = \operatorname{Im} \alpha_0$ determines the non-radiative decay rate of the guided mode
%(small imaginary contribution to $\Omega$)
%
\begin{equation}\label{eq:Gamma}
    \Gamma(k) = \alpha''_0 \beta(k)  \,, \quad \beta(k) =-\frac{\partial  \Omega(k)}{\partial \alpha'_0} = \frac{\Omega(k)-v(k) k}{\alpha'_0} \,,
\end{equation}
where $v(k) = \partial \Omega / \partial k$ is the group velocity.

\begin{figure}[t]
    \centering
    \includegraphics[width=0.85\linewidth]{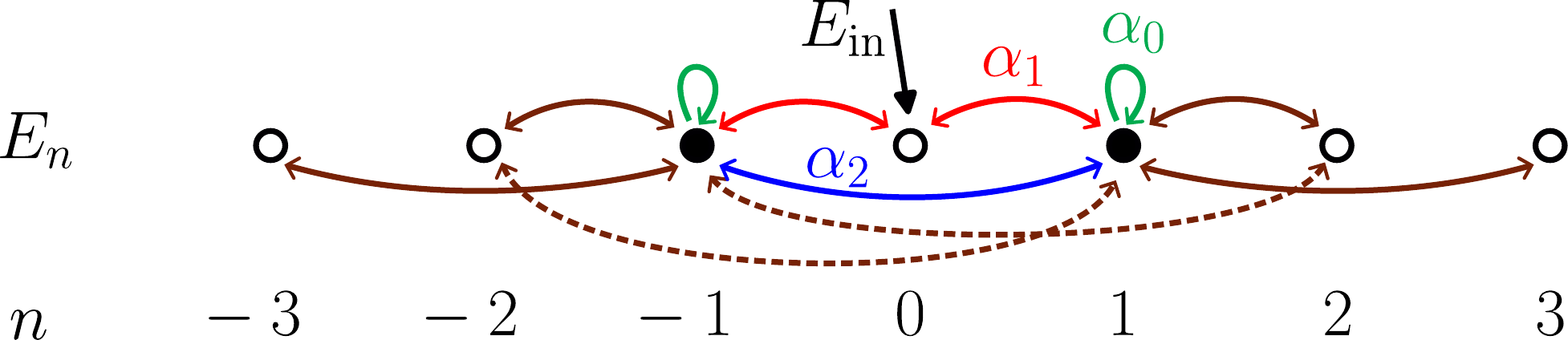}
    \caption{Sketch of the major processes and the corresponding coupling coefficients $\alpha_n$ determining the widths and spectral positions of resonances in low-contrast meta-waveguides. }
    \label{fig:scheme-int}
\end{figure}

We focus on the lowest frequency resonances, where the harmonics $E_{\pm1}$ are enhanced, and on small incidence angles $\theta \ll 1$. In this case, the resonance condition is given by $\omega \approx \Omega(g)$, and the corresponding truncated set of equations derived from Eq.~\eqref{eq:En} takes the form
\begin{align}\label{eq:system-trunc}
    E_0&=2\pi \rmi \omega\left(\alpha_0 E_0+\alpha_{-1} E_{1}+\alpha_{1}E_{-1}\right)+E_{\rm in}\;,\\ \nonumber
    \begin{split}
    E_{\pm1}&=\frac{2\pi \omega^2}{\varkappa_{\pm1}}\bigg[\alpha_0 E_{\pm 1}+ \alpha_{\pm1} E_0+ \alpha_{\pm2} E_{\mp1}\\
    &\qquad\qquad\qquad\qquad\qquad+\sum_{|n|\geq 2}\alpha_{-n\pm1}E_{n}\bigg]\,,\end{split}\\ \nonumber
    E_{n}&=\frac{2\pi \omega^2}{\varkappa_{n}}\left(\alpha_0 E_{n}+\alpha_{n-1}E_{1}+\alpha_{n+1}E_{-1}\right)\,.
\end{align}
This system captures the main processes that determine the spectral positions and linewidths of the resonances in low-contrast MWs, as illustrated in Fig.~\ref{fig:scheme-int}. The incident plane wave resonantly excites the modes with $n = \pm1$, which, in turn, interact with each other and with higher-order modes. The key parameters of the system are $\alpha_0$, $\alpha_{\pm1}$, and $\alpha_{\pm2}$, which govern, respectivley, the resonance frequencies, the coupling between the incident light and the resonant modes, and the direct interaction between the resonant modes.

The system of Eqs.~\eqref{eq:system-trunc} is transformed into the equations for the amplitudes $E_0$ and $E_n$ ($|n| \geq 2$)
\begin{align}\label{eq:E0}
    E_0 &= 2 \pi \rmi \omega t_0 \left(\alpha_{-1} E_{1}+\alpha_{1}E_{-1}\right) + t_0 E_{\rm in} \,, \\ \nonumber
    E_n &= \frac{2\pi \omega^2}{\varkappa_n - 2\pi \omega^2 \alpha_0} \left(\alpha_{n-1} E_{1}+\alpha_{n+1}E_{-1}\right) \,,
\end{align}
and the pair of coupled equations for the amplitudes $E_{\pm 1}$
\begin{equation}\label{eq:Epm}
    \frac{1}{2\pi\omega^2}(\varkappa_{\pm1}-2\pi\omega^2\tilde{\alpha}_{\pm0})E_{\pm1}-\tilde{\alpha}_{\pm2} E_{\mp1}=\alpha_{\pm1} t_0 E_{\rm in}\,.
\end{equation}
Here, $t_0 = 1/(1-2\pi\rmi\omega\alpha_0)$ is the amplitude transmission coefficient of normally incident light through the uniform slab, and $\tilde{\alpha}_{\pm0}$ and $\tilde{\alpha}_{\pm2}$ are the coupling parameters renormalized by higher-order interactions
\begin{align}\label{eq:renorm-a}
    \tilde{\alpha}_{\pm0}&=\alpha_0+2\pi \rmi\omega t_0\alpha_{\pm1} \alpha_{\mp1}+\sum_{|n|\geq 2}\frac{2\pi \omega^2\alpha_{-n\pm1}\alpha_{n\mp1}}{\varkappa_{n}-2\pi \omega^2\alpha_0}\,,\\\nonumber 
    \tilde{\alpha}_{\pm2}&=\alpha_{\pm2}+2\pi \rmi\omega t_0\alpha_{\pm1}^2+\sum_{|n|\geq 2}\frac{2\pi \omega^2\alpha_{-n\pm1}\alpha_{n\pm1}}{\varkappa_{n}-2\pi \omega^2\alpha_0}\,.
\end{align}
In the following, we analyze the resonance contributions to the near-field amplitude for both normal and oblique incidence of radiation, and for symmetric and asymmetric gratings.

\subsection{Normal incidence of light on MW with symmetric grating}\label{Sec_IIA}

For symmetric grating with $\alpha (x) = \alpha(-x)$, the Fourier harmonics satisfy $\alpha_n = \alpha_{-n}$. Furthermore, under normal incidence of radiation ($\theta = 0$), we have $\varkappa_{+1} = \varkappa_{-1}$ and $\tilde{\alpha}_n = \tilde{\alpha}_{-n}$. Consequently, the amplitudes $E_1$ and $E_{-1}$ coincide, i.e., the incident light excites the symmetric standing MW wave. This is the leaky MW mode that is coupled to the incident radiation. 

The field amplitudes given by Eqs~\eqref{eq:E0} and~\eqref{eq:Epm} assume the form
\begin{align}\label{eq:E-normal-sym}
    E_0 = t_0 E_{\rm in} + 4 \pi \rmi \omega \alpha_1 t_0 E_1  \,, \\ \nonumber
   E_{\pm 1} =  \frac{2\pi \omega^2 \alpha_1 t_0 E_{\rm in}}{\varkappa_1 - 2\pi\omega^2 (\tilde{\alpha}_0 + \tilde{\alpha}_2)} \,. 
\end{align}
At the resonance, the amplitudes have the pole structure,
\begin{equation}\label{eq:pole}
   \frac{2\pi \omega^2}{\varkappa_{1}-2\pi\omega^2(\tilde{\alpha}_0+\tilde{\alpha}_2)}
       \approx
       \frac{- \beta}{\omega - \omega_{\rm s} + \rmi (\Gamma_0 + \Gamma)} \;,
\end{equation}
where $\beta = \beta(g) = 4 \pi^2 \Omega^3 \alpha'_0 /(1+8\pi^2 \Omega^2 \alpha_0'^2)$,
$\Omega = \Omega(g)$, $\omega_{\rm s}$ is the resonant frequency, and $\Gamma_0$ and $\Gamma = \alpha''_0 \beta$ are the radiative and non-radiative broadening, respectively.
To the second order in the permittivity modulation, $\omega_{\rm s}$ and $\Gamma_0$ have the form
\begin{align}\label{eq:omega_s}
&\omega_{\rm s}  = \Omega - \alpha_2 \beta+\bigg(3-\frac{\beta}{(2\pi)^4\Omega^5\alpha_0'^3}\bigg)\frac{\alpha_2^2 \beta^2}{2\Omega} 
\\ \nonumber
&+ 4 \pi \Omega \alpha_1^2 \beta  \operatorname{Im} t_0- 2 \pi \Omega^2 \beta \sum_{n \geq 2} \frac{(\alpha_{n-1} + \alpha_{n+1})^2}{\varkappa_{n}- 
\varkappa_{1}} \,
\end{align}
and
\begin{equation}\label{eq:Gamma0}
\Gamma_0 = 4 \pi \Omega \alpha_1^2 \beta \operatorname{Re} t_0  \,.
\end{equation}
The resonant frequency $\omega_{\rm s}$ is slightly shifted from $\Omega$. The dominant contribution to the shift is proportional to the coefficient $\alpha_2$ in the Fourier series~\eqref{eq:alpha}, which describes the direct coupling of the $n = \pm 1$ modes. This coupling leads to the spectral splitting between the leaky symmetric standing waves
with $E_{+1} = E_{-1}$ and optically uncoupled (BIC)
antisymmetric standing waves with $E_{+1} = - E_{-1}$~\cite{Kazarinov1976}.
Note, that the latter has the resonant frequency
\begin{align}\label{eq:omega_a}
\omega_{\rm a}  = \Omega + \alpha_2 \beta + &\bigg(3-\frac{\beta}{(2\pi)^4\Omega^5\alpha_0'^3}\bigg) \frac{\alpha_2^2 \beta^2}{2\Omega}   
\\ \nonumber
&- 2 \pi \Omega^2 \beta \sum_{n \geq 2} \frac{(\alpha_{n-1}- \alpha_{n+1})^2}{\varkappa_{n}- 
\varkappa_{1}} \,
\end{align}
and vanishing radiative broadening at the normal incidence of light. 

Thus, the spectral dependence of the field amplitudes at the resonance can be presented in the general form 
\begin{align}\label{eq:E-normal-sym-final}
    & E_0 = \left[ t_0 - \frac{\rmi \Gamma_0}{\omega - \omega_{\rm s} +\rmi (\Gamma_0 + \Gamma)} \frac{t_0}{t_0^*}  \right] E_{\rm in} \,, \\  \nonumber
    & E_{\pm 1} = -\sqrt{\frac{\beta}{4\pi \Omega^2}}\frac{\sqrt{\Omega \Gamma_0}}{\omega - \omega_{\rm s} +\rmi (\Gamma_0 + \Gamma)}\frac{t_0}{|t_0|}  E_{\rm in} \,,
    % & E_{\pm 1} = \frac{- \alpha_1 \beta t_0}{\omega - \omega_{\rm r} +\rmi (\Gamma_0 + \Gamma)}  E_{\rm in} \,,\\
    % & E_{\pm 1} = -\sqrt{\frac{\beta \Gamma_0}{4\pi \Omega}}\frac{1}{\omega - \omega_{\rm r} +\rmi (\Gamma_0 + \Gamma)}\frac{t_0}{t_0^*}  E_{\rm in} \,,\\
    % & E_{\pm 1} = -\sqrt{\frac{v \varkappa_1}{2g}}\frac{\sqrt{\Omega\Gamma_0}}{\omega - \omega_{\rm r} +\rmi (\Gamma_0 + \Gamma)}\frac{t_0}{t_0^*}  E_{\rm in} \,,
\end{align}
Here, we took into account that $\operatorname{Re} t_0 = |t_0|^2$ for any transmission $t_0$ and reflection $r_0$ coefficients satisfying the relations $t_0 = 1 + r_0$ and 
$|t_0|^2 + |r_0|^2 = 1$.

\subsection{Oblique incidence of light on MW with symmetric grating}\label{Sec_IIB}

At oblique incidence of light with the small incidence angle $\theta$,
$\varkappa_{\pm1}(\omega,\theta) \approx \varkappa_{1,0}\pm(g \omega /\varkappa_{1,0}) \theta$, where 
$\varkappa_{1,0} = \varkappa_{\pm1}(\omega, 0) = \sqrt{g^2 - \omega^2}$. 
Then, the system of Eqs.~\eqref{eq:Epm} written in the basis of symmetric and antisymmetric states $E_{\rm s/a}=(E_1\pm E_{-1})/2$ at the MW resonance assumes the matrix form
\begin{equation}
    \begin{pmatrix}\label{eq:Epm-eqs}
    \omega - \varepsilon_{\rm s}  & - \delta_{\theta} \\
        -\delta_{\theta} & \omega - \varepsilon_{\rm a}
    \end{pmatrix}
    \begin{pmatrix}
        E_{\rm s}\\
        E_{\rm a}
    \end{pmatrix}=-    \begin{pmatrix}
        \beta \alpha_1 t_0 E_{\rm in}\\
        0
    \end{pmatrix},
\end{equation}
where $\varepsilon_{\rm s} = \omega_{\rm s} - \rmi (\Gamma_0 + \Gamma)$ and 
$\varepsilon_{\rm a} = \omega_{\rm a} - \rmi \Gamma$ are the complex eigenfrequencies of the symmetric and antisymmetric modes at the normal incidence of light, respectively. Here, $\omega_{\rm s}$, $\Gamma_0$, and $\omega_{\rm a}$ are given by Eqs.~\eqref{eq:omega_s}–\eqref{eq:omega_a}, and 
\begin{equation}
    \delta_{\theta} = v(g) \Omega \theta = 
    \frac{g\theta}{1+ 8 \pi^2 \Omega^2 \alpha_0'^2} 
\end{equation}
is the frequency shift induced by the oblique incidence. 

Equation~\eqref{eq:Epm-eqs} shows that the optical response contains two poles, whose positions are determined by the eigenfrequencies
\begin{equation}
    \varepsilon_{\rm 1,2} = \frac{\omega_{\rm s}+\omega_{\rm a} 
    -\rmi (\Gamma_0 + 2 \Gamma)}{2}  \pm \sqrt{\left(\frac{\omega_{\rm s}-\omega_{\rm a} - \rmi \Gamma_0}{2}\right)^2+\delta_\theta^2} \,.
\end{equation}

At very small angles of incidence, 
when $|\delta_\theta |\ll |\omega_{\rm s} - \omega_{\rm a}|\approx 2|\alpha_2| \beta $,
%the poles are located at 
the eigenfrequencies can be approximated as
\begin{align}
    &\varepsilon_1=\omega_{\rm s}-\frac{\delta_{\theta}^2}{2\alpha_2 \beta}-\rmi (\Gamma_0+\Gamma) \,,\\
    &\varepsilon_2=\omega_{\rm a}+\frac{\delta^2_\theta}{2\alpha_2 \beta}-\rmi\left[\Gamma_0\bigg(\frac{\delta_\theta}{2\alpha_2 \beta}\bigg)^2+\Gamma\right] \,.
\end{align}
Then, the spectral dependence of the field amplitudes 
$E_{\pm 1}$ has the form
\begin{align}\label{eq:Epm_oblique1}
    E_{\pm1} = &-\sqrt{\frac{\beta}{4\pi \Omega^2}}
    \bigg[
    \frac{\sqrt{\Omega\Gamma_0}}{\omega-\omega_1 + \rmi [\Gamma_0 + \Gamma]}
    \\ \nonumber & \pm
    \frac{\sqrt{\Omega\Gamma_0} (\nu/ 2)}{\omega-\omega_2 + \rmi [\Gamma_0 (\nu /2)^2  + \Gamma]}
    \bigg]
    \frac{t_0}{|t_0|}E_{\rm in} \,,
\end{align}
where $\omega_{1,2} = \operatorname{Re} \varepsilon_{1,2}$ and $\nu = \delta_\theta /(\alpha_2 \beta)$. 
Here, $|\nu| \ll 1$.
The poles correspond to the resonant excitation of the symmetric (leaky) and antisymmetric (now quasi-BIC) modes, respectively~\cite{Koshelev2021}. 
The quasi-BIC mode appears in the optical response due to its coupling with the leaky mode at $\theta \neq 0$. Its radiative broadening $\Gamma_0 (\nu /2)^2$ is, therefore, much smaller than that of the leaky mode $\Gamma_0$.

In the opposite case of relatively large angles of incidence, 
$|\delta_\theta| \gg |\alpha_2 \beta|$, $|\nu| \gg 1$, the modes with $n=+1$ and $n=-1$ get spectrally separated by $2\delta_\theta$. The corresponding pole contributions to the field amplitudes assume the form
\begin{align}\label{eq:Epm_oblique2}
    E_{\pm1} = &-\sqrt{\frac{\beta}{4\pi \Omega^2}}
    \bigg[
    \frac{\sqrt{\Omega\Gamma_0}}{\omega-\omega_{1,2} + \rmi (\Gamma_0/2 + \Gamma)}
    \\ \nonumber & +\frac{1}{2\nu}
    \frac{\sqrt{\Omega\Gamma_0}}{\omega-\omega_{2,1} + \rmi (\Gamma_0/2  + \Gamma)}
    \bigg]
    \frac{t_0}{|t_0|}E_{\rm in} 
\end{align}
with $\omega_{1,2}=(\omega_{\rm s}+
\omega_{\rm a})/2 \pm \delta_\theta$. The radiative broadening of each mode is given by $\Gamma_0/2$.

\subsection{Normal incidence of light on asymmetric MW}
\label{sec:IIC}

For a MW with asymmetric grating, the Fourier coefficients $\alpha_n$ and $\alpha_{-n}$ do not necessarily coincide, although they remain related by complex conjugation. 
By an appropriate choice of the coordinate origin, we can set $\alpha_1 = \alpha_{-1}$ and express the permittivity as
\begin{equation}
\alpha(x) = \alpha_0 + 2 \alpha_1 \cos{(gx)} 
+ \sum_{n \geq 2} 2 |\alpha_n| \cos{(ngx + \varphi_n)} \,,
\end{equation}
where $\alpha_n = |\alpha_n| \exp(\rmi \varphi_n)$.
Then, for the normal incidence of light, $\tilde{\alpha}_{+0} = \tilde{\alpha}_{-0}$ 
and also $\varkappa_{+1} = \varkappa_{-1}$.

As it follows from Eqs.~\eqref{eq:Epm}, the eigenfrequencies are now determined by the equation
\begin{equation}
    (\varkappa_1 - 2\pi \omega^2 \tilde{\alpha}_0)^2 - (2\pi \omega^2)^2 
    \tilde{\alpha}_{+2}  \tilde{\alpha}_{-2} = 0 \,,
\end{equation}
which yields 
\begin{align}
    &\varepsilon_{1,2} = \Omega \mp |\alpha_2| \beta-\rmi \Gamma -2\pi \Omega^2\beta\sum_{n\geq 2}\frac{|\alpha_{n+1}\pm\alpha_{n-1}\e^{\rmi\varphi_2}|^2}{\varkappa_n-\varkappa_1} \nonumber \\ 
    & -2\pi\rmi \Omega \alpha_1^2\beta t_0 (1\pm\cos \varphi_2)
    +\bigg(3-\frac{\beta}{(2\pi)^4\Omega^5\alpha_0'^3}\bigg)
    \frac{|\alpha_2|^2 \beta^2}{2\Omega}\,.
\end{align}
The corresponding pole contributions to the field amplitudes have the form
\begin{align}
    E_{\pm1} = & -\sqrt{\frac{\beta}{4\pi \Omega^2}}
    \bigg[
    \frac{\sqrt{\Omega\Gamma_0}\cos(\varphi_2/2)}{\omega-\omega_1+\rmi[\Gamma_0 \cos^2 (\varphi_2/2) +\Gamma]}    \\ \nonumber
    & \mp \rmi
    \frac{\sqrt{\Omega\Gamma_0}\sin(\varphi_2/2)}{\omega-\omega_2 + \rmi [\Gamma_0 \sin^2 (\varphi_2/2) +\Gamma]}
    \bigg]
    \frac{t_0 \e^{\pm\rmi\varphi_2/2}}{|t_0|}E_{\rm in} \,
\end{align}
with $\omega_{1,2} = \operatorname{Re} \varepsilon_{1,2}$. The radiative broadening is redistributed between the two modes with the partition factors
$\cos^2 (\varphi_2/2)$ and $\sin^2 (\varphi_2/2)$.

\subsection{Second harmonic radiation}

Having calculated the near field at the fundamental frequency, we now proceed to the evaluation of the polarization and emission at the double frequency. The polarization $\bm P^{(2)}$  is given by Eq.~\eqref{eq:P2}, which yields
\begin{align}
    P^{(2)}_x &= \sum_{n,m} (\chi_{xyy} + \rmi ng \chi') E_{n} E_m  
    \e^{\rmi [2 q_x + (n+m) g] x}  \,, \\ \nonumber
    P^{(2)}_y & = \chi_{yyy} \sum_{n,m} E_{n} E_m  
    \e^{\rmi [2 q_x + (n+m) g] x} \,.
\end{align}

The second harmonic field $\bm E^{(2)}$ is then found from the wave equation 
\begin{equation}\label{eq:2w-eq}
    \operatorname{rot}\operatorname{rot} \bm E^{(2)} - (2\omega)^2\bm E^{(2)} = 
    4\pi (2\omega)^2[\alpha(x)\bm E^{(2)}_\parallel+\bm P^{(2)}]\delta(z) 
\end{equation}
with the source term $\propto\bm P^{(2)}\delta(z)$.
We solve the equation by expanding the field into the spatial Fourier harmonics in a manner similar to the case of linear response.

Near MW resonances, the local field is determined primarily by the harmonics $E_{\pm1}$. Therefore, the dominant contribution to the forward transmitted 
%and mirror reflected beams
beam at $2\omega$ arises from the product $E_{+1}E_{-1}$. The amplitude of the emitted second harmonic field is then given by
\begin{equation}\label{eq:Ej_2w}
    E_{j}^{(2)}=8\pi \rmi\omega  t_0(2\omega)\chi_{jyy}E_{+1}E_{-1}
\end{equation}
provided the second harmonic radiation does not fall (occasionally) in a higher frequency MW resonance. Note that the gradient term $\propto \chi'$ does not contribute to the this field.
The polarization of the emitted radiation is determined by the tensor $\chi$. The spectral dependence of the intensity $I^{(2)}=|\bm E^{(2)}|^2/(2\pi)$ contains two resonances at the frequencies $\omega_{1,2}$, which follows from the frequency dependence of the amplitudes $E_{\pm1}$.

The diffracted beams ''$\pm1_{2\omega}$'' are mainly determined by the terms $E_0E_{\pm1}$, $E_{\pm2}E_{\mp1}$, $E_{+1}E_{-1}$, and $E_{\pm1}^2$. Whereas the first two terms generate the polarization $P^{(2)} \propto \exp[\rmi(2q_x\pm g)x]$ which directly emits the beams ''$\pm1_{2\omega}$'', the other two terms contribute to the diffraction beams via additional scattering by the MW grating. 

\section{Results and Discussion}\label{sec:res}

\begin{figure}[t]
    \centering
    \includegraphics[width=1\linewidth]{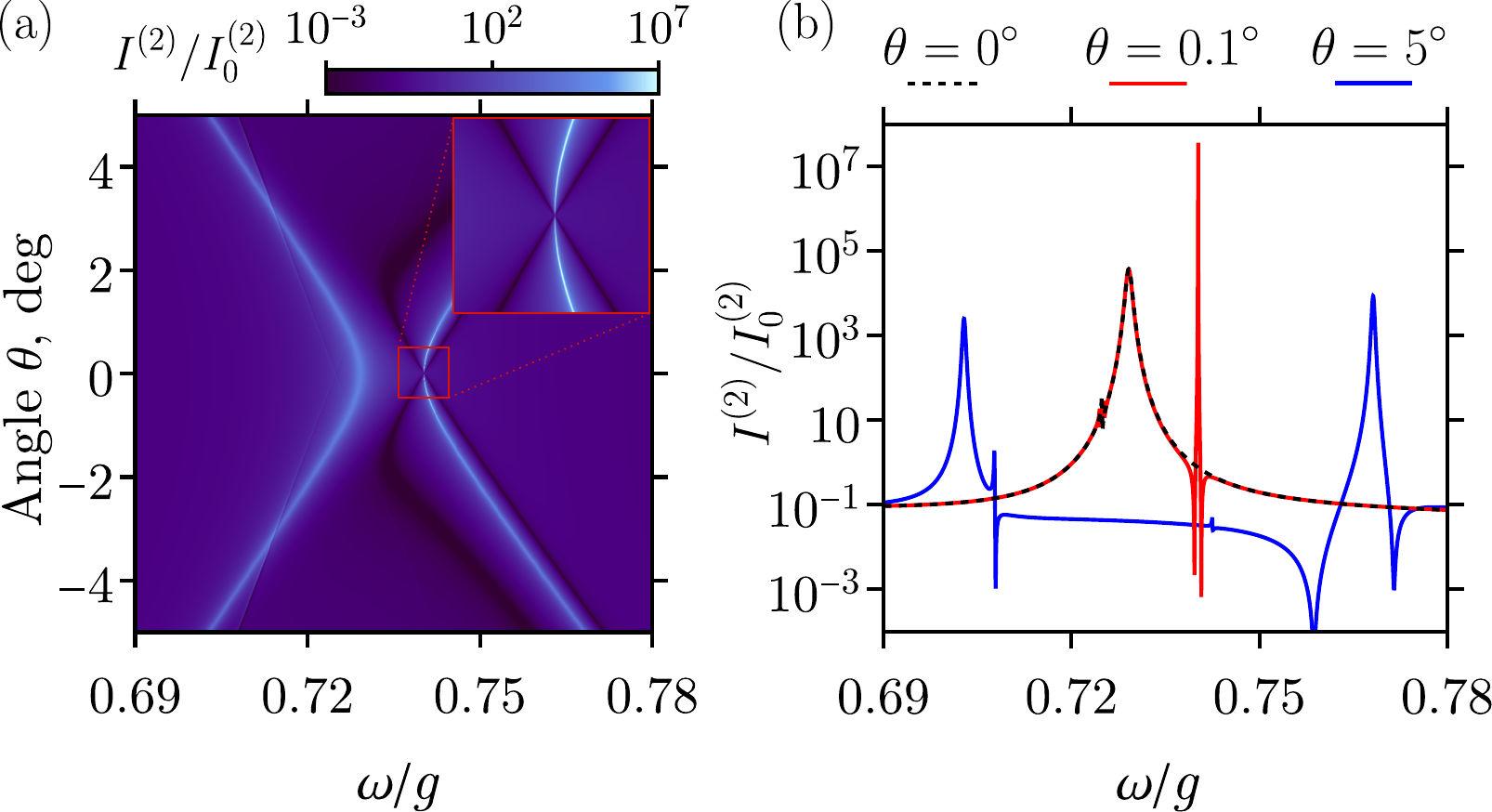}
    \caption{Resonant enhancement of SHG by the meta-waveguide. (a) SHG enhancement as a function of the light frequency and the angle of incidence. (b) Spectral dependence of the SHG enhancement for three different angles of incidence. The figures are calculated for the symmetric MW with the parameters: $\alpha_0 g=0.2$, $\alpha_1 g=0.01$, 
    $\alpha_2 g=0.005$, and the nonlinear susceptibility $\chi_{yyy}$.}
\label{fig:SHG-Intens}
\end{figure}

Now we present the results of numerical and analytical calculations of SHG, compare them, 
and discuss the enhancement of SHG at the meta-waveguide resonances.

Figure~\ref{fig:SHG-Intens}~(a) shows the intensity of the forward emitted second harmonic radiation $I^{(2)}$ as a function of the fundamental frequency and the angle of incidence. 
The intensity $I^{(2)}$ is normalized to the reference value
$I_0^{(2)} = 2\omega^2 \chi^2 E_0^4$,
corresponding to the SHG in the absence of the MW. The data are obtained by numerically solving Eqs.~\eqref{eq:En} and~\eqref{eq:2w-eq} for a symmetric MW with the parameters given in the figure caption. 
The figure reveals a pronounced enhancement of the SHG signal at the MW resonances, with the enhancement factor up to $10^7$.

At the normal incidence of radiation [$\theta = 0$, see also the black dashed curve in Fig.~\ref{fig:SHG-Intens}~(b)], the SHG signal exhibits a single resonance. This resonance is associated 
with the excitation of the symmetric standing wave in the MW, see Sec.~\ref{Sec_IIA} for details. A slight deviation of the light incidence from the normal [$\theta = 0.1^\circ$ in Fig.~\ref{fig:SHG-Intens}~(b)] leads to the excitation of the antisymmetric quasi-BIC mode, see Sec.~\ref{Sec_IIB}.
This gives rise to an additional, exceptionally narrow and high, peak in the SHG excitation spectrum.   
With a further increase of the incidence angle [$\theta = 5^\circ$ in Fig.~\ref{fig:SHG-Intens}~(b)], the two peaks get spectrally separated and acquire comparable radiative broadening. 
Note that the calculated SHG spectrum contains additional weak resonances (e.g., a small feature on the left wing of the black  curve and a narrow feature at the low-energy peak of the blue curve) originating from the occasional resonance of the second harmonic field with the MW mode with the wave vector $3g$.

\begin{figure}[t]
    \centering
    \includegraphics[width=1\linewidth]{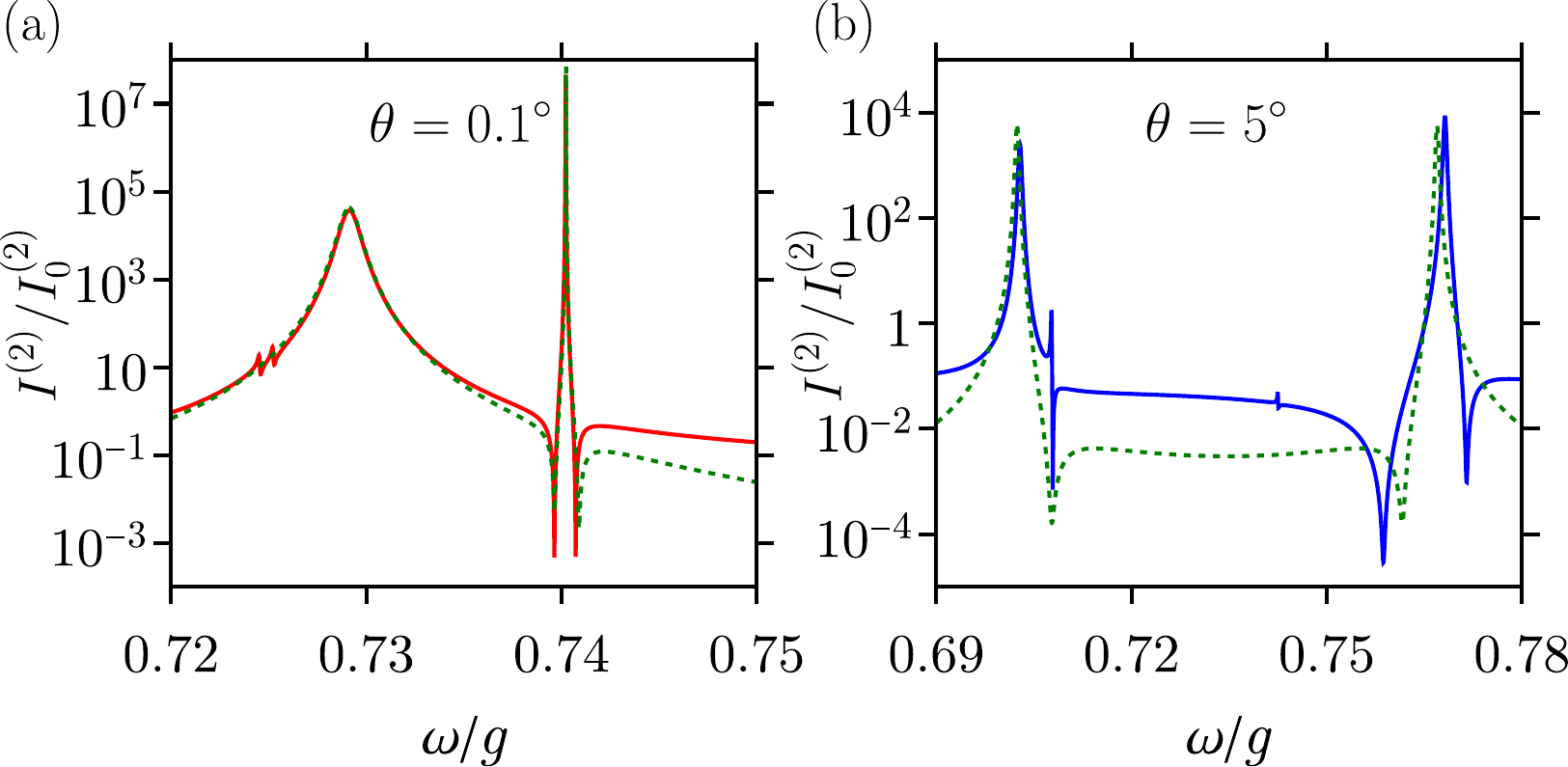}
    \caption{Spectral dependence of the SHG enhancement calculated numerically (solid curves) and plotted following analytical equations (dashed curves) for two different angles of incidence.  }
    \label{fig:SHG-Analytic}
\end{figure}

The resonant contributions to the SHG signal are well captured by the analytical theory developed above. Figure~\ref{fig:SHG-Analytic} compares the numerically calculated (solid curves) and analytically derived (dashed curves) spectral dependencies of the SHG intensity.
The analytical dependencies are plotted for the emitted field Eq.~\eqref{eq:Ej_2w} with $E_{\pm 1}$ given by Eq.~\eqref{eq:Epm_oblique1} for small angles of incidence [Fig.~\ref{fig:SHG-Analytic}(a)] and Eq.~\eqref{eq:Epm_oblique2} for larger angles of incidence  
[Fig.~\ref{fig:SHG-Analytic}(b)], respectively. The figure reveals that the analytical theory describes well both the magnitudes and the widths of the resonances. 

At the resonances associated with the excitation of the leaky or quasi-BIC modes at small angles of incidence [Fig.~\ref{fig:SHG-Analytic}(a)], the resonant contributions are given by 
\begin{eqnarray}\label{eq:Int2W}
   && I^{(2)}_{\rm sym} = \ \frac{\beta^2 |t_0(2\omega)|^2}{4 \pi^2 \Omega^4}
    \frac{\Omega^2 \Gamma_0^2}{[(\omega - \omega_1)^2 + (\Gamma + \Gamma_0)^2]^2} I_0^{(2)} 
    , \\
   &&  I^{(2)}_{\rm BIC}  = \frac{\beta^2 |t_0(2\omega)|^2}{4 \pi^2 \Omega^4}
    \frac{\Omega^2 (\nu /2)^4 \Gamma_0^2 }{\{[\omega - \omega_2]^2 + [\Gamma + (\nu /2)^2 \Gamma_0]^2\}^2} I_0^{(2)} , \nonumber
\end{eqnarray}
where $|\nu| = |\delta_\theta / (\alpha_2 \beta)| \ll 1$.
Note that Eqs.~\eqref{eq:Int2W} are also valid for the normal incidence of light on an asymmetric MW with a small degree of asymmetry, in this case $\nu=\phi_2$ is the asymmetry parameter, see Sec.~\ref{sec:IIC}.
%$\nu=\varphi_2\ll1$.   
In the MW with negligible non-radiative broadening 
$\Gamma$, the SHG enhancement at the leaky mode resonance at $\Omega \alpha_0 \sim 1$ can be estimated as $I^{(2)}/ I^{(2)}_0 \sim (\Omega/\Gamma_0)^2 \sim (\Re \alpha_0 /\alpha_1)^4$. The SHG enhancement at the quasi-BIC resonance $I^{(2)}/ I^{(2)}_0 \sim (\Omega/\Gamma_0)^2 (1/\nu)^4$ is even stronger by the factor $(1/\nu)^4 \sim (\alpha_2/\Re\alpha_0)^4 (1/\theta)^4$ at $\nu \ll 1$ and scales as $(1/\theta)^4$ with the angle of incidence. The dramatic growth of $I^{(2)}$ at very small angles is limited by non-radiative broadening $\Gamma$. Equation~\eqref{eq:Int2W} reveals that the maximal enhancement 
$I^{(2)}/I^{(2)}_0 \sim(\Omega/\Gamma)^4 \sim(\Re \alpha_0/\Im \alpha_0)^4$ 
is achieved at the incidence angle
\begin{equation}
    \theta = \frac{4 \pi^{3/2} \Omega^{5/2} \alpha_2 \Re \alpha_0 \sqrt{\Im \alpha_0}}{g \alpha_1 \sqrt{\operatorname{Re} t_0}} \sim \sqrt{\frac{\Im\alpha_0}{\Re\alpha_0}} \,.
\end{equation}

At larger angles of incidence, when $|\nu| \gg 1$ and the resonances get independent [Fig.~\ref{fig:SHG-Analytic}(b)], the intensity of the second harmonic radiation at each of the resonances has the form
\begin{equation}
     I^{(2)} = \frac{\beta^2 |t_0(2\omega)|^2}{16 \pi^2 \Omega^4\nu^2 }
    \frac{\Omega^2 \Gamma_0^2}{[(\omega - \omega_{1,2})^2 + 
    (\Gamma + \Gamma_0/2)^2]^2} I_0^{(2)} \,.
\end{equation}
The resonances have the radiative width $\Gamma_0/2$ and the magnitude decreased by the factor 
$1/\nu^2$.

\begin{figure}[t]
    \centering
    \includegraphics[width=1\linewidth]{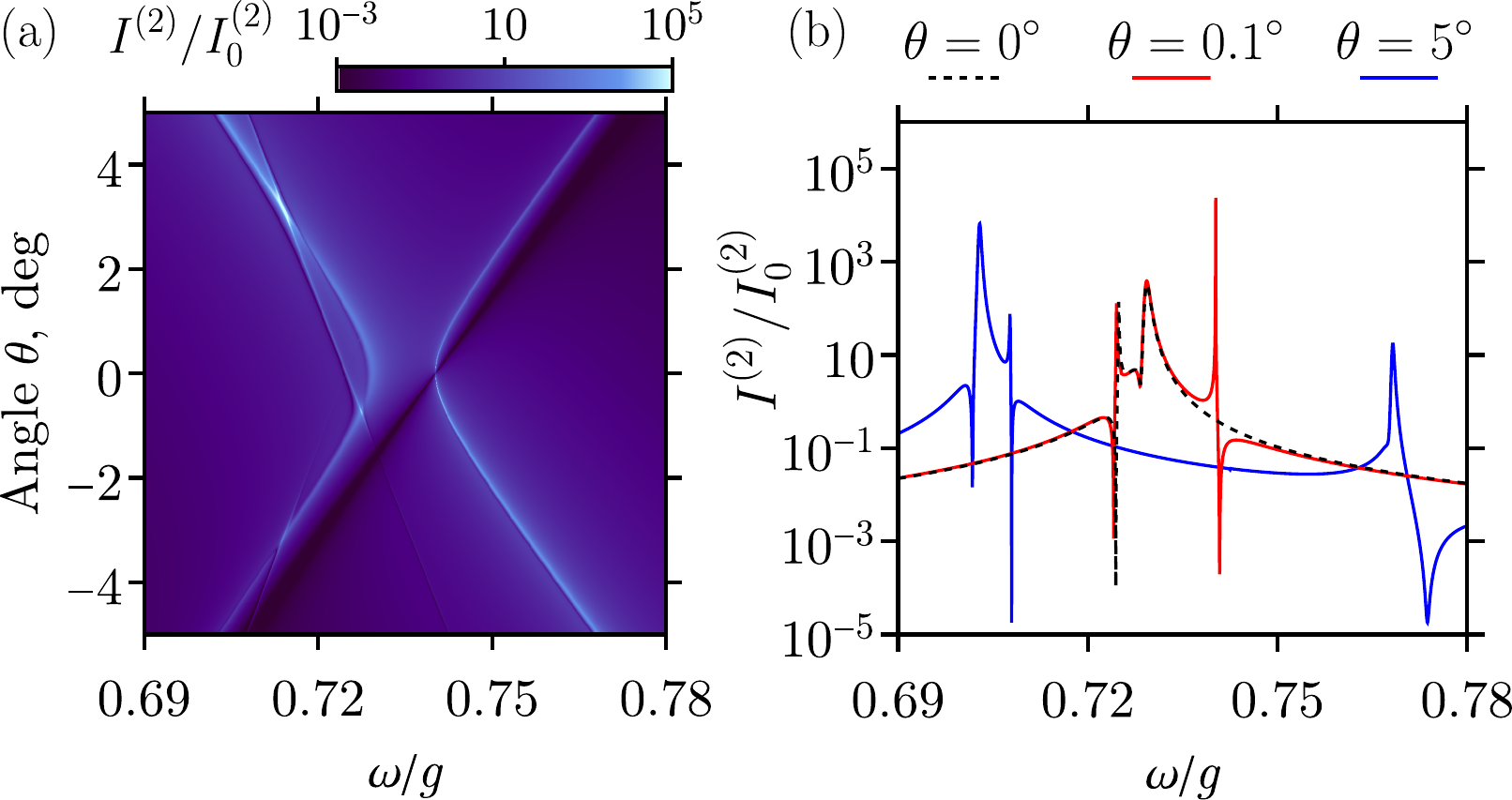}
    \caption{(a) Intensity of the second harmonic diffracted beam ''$-1_{2\omega}$'' as a function of the frequency $\omega$ and the angle of incidence $\theta$. (b) Spectral dependence of the intensity for three different angles of incidence. The figures are calculated for the same parameters as Fig.~\ref{fig:SHG-Intens}.}
    \label{fig:SHG-Difr-Map}
\end{figure}

Figure~\ref{fig:SHG-Difr-Map} shows the intensity of the second harmonic diffracted beam ''$-1_{2\omega}$'' (see Fig.~\ref{fig:scheme}) as a function of the frequency $\omega$ and the angle $\theta$ of the incident beam. We remind that, at the normal incidence of radiation, 
the diffracted beams at $2\omega$ emerge when $g < 2 \omega/c$, where $g$ is the MW wave vector, whereas the diffraction of the fundamental beam occurs for $g < \omega/c$. Therefore, in the range $\omega/c < g < 2\omega/c$, 
the diffraction picture contains only the second harmonic beams.
Figure ~\ref{fig:SHG-Difr-Map} reveals that the spectral dependence of the diffraction intensity contains the resonances associated with the excitation of guided modes, similar to those observed for the forward emitted second harmonic radiation.

\begin{figure}[t]
    \centering
    \includegraphics[width=1\linewidth]{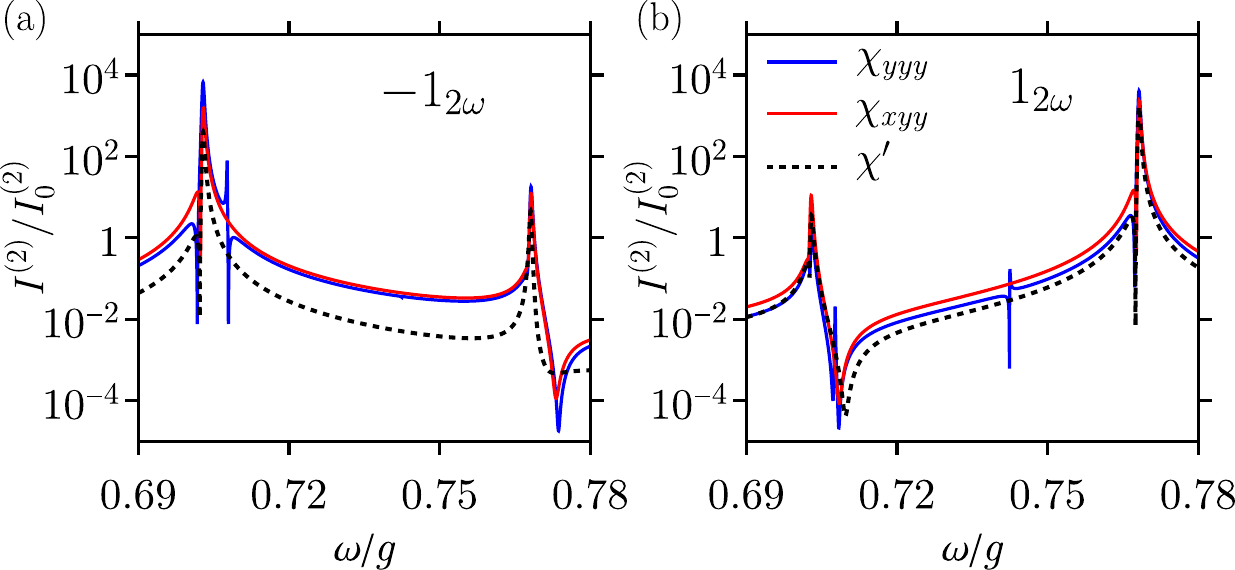}
    \caption{Excitation spectra of the second harmonic diffracted beams ''$-1_{2\omega}$'' and ''$+1_{2\omega}$'' for different sources of SHG in the non-linear crystal. Blue and red curves correspond to the $\chi_{yyy}$ and $\chi_{xyy}$ components of the second-order nonlinear susceptibility, respectively. Black dashes curves correspond to SHG due to the spatial structure of the field, which does not require space inversion asymmetry in the crystal. The curves are calculated for the parameters given in the caption of Fig.~\ref{fig:SHG-Intens}, $\theta = 5^\circ$, and equal 
    $\chi_{yyy}$, $\chi_{xyy}$, and $\chi' g$.}
    \label{fig:SHG-Difr-Int}
 \end{figure}

Interestingly, the meta-waveguide enables the generation of the polarization at double frequency in the 2D nonlinear crystal and the second harmonic diffracted beams even if both the meta-waveguide and the 2D crystal are centrosymmetric. The corresponding mechanism of SHG originates from the strong in-plane inhomogeneity of the near field, as characterized by the parameter $\chi'$ in Eq.~\eqref{eq:P2}.
Figure~\ref{fig:SHG-Difr-Int} shows the excitation spectra of the second harmonic diffracted beams ''$+1_{2\omega}$'' and ''$-1_{2\omega}$'' for different sources of SHG in the nonlinear crystal, including the standard second-order nonlinearity and the nonlinearity due to the field inhomogeneity. All spectra contain the MW resonances. 
The spectrum calculated for the $\chi_{yyy}$ component of the second-order nonlinear susceptibility also displays additional features arising from the resonances of the second harmonic field with high MW modes. 
The second harmonic diffraction is strongly directional: depending on the frequency $\omega$ either ''$-1_{2\omega}$'' or ''$+1_{2\omega}$'' beam predominates. At the resonances, the intensity of the diffracted beams can be estimated as $\sim(\Omega/\Gamma_0)I_{0}^{(2)}$ and $\sim\nu^{-2}(\Omega/\Gamma_0)I_{0}^{(2)}$ with $\nu=\delta_\theta /(\alpha_2 \beta)\gg1$. 

\begin{figure}[t]
    \centering
    \includegraphics[width=1\linewidth]{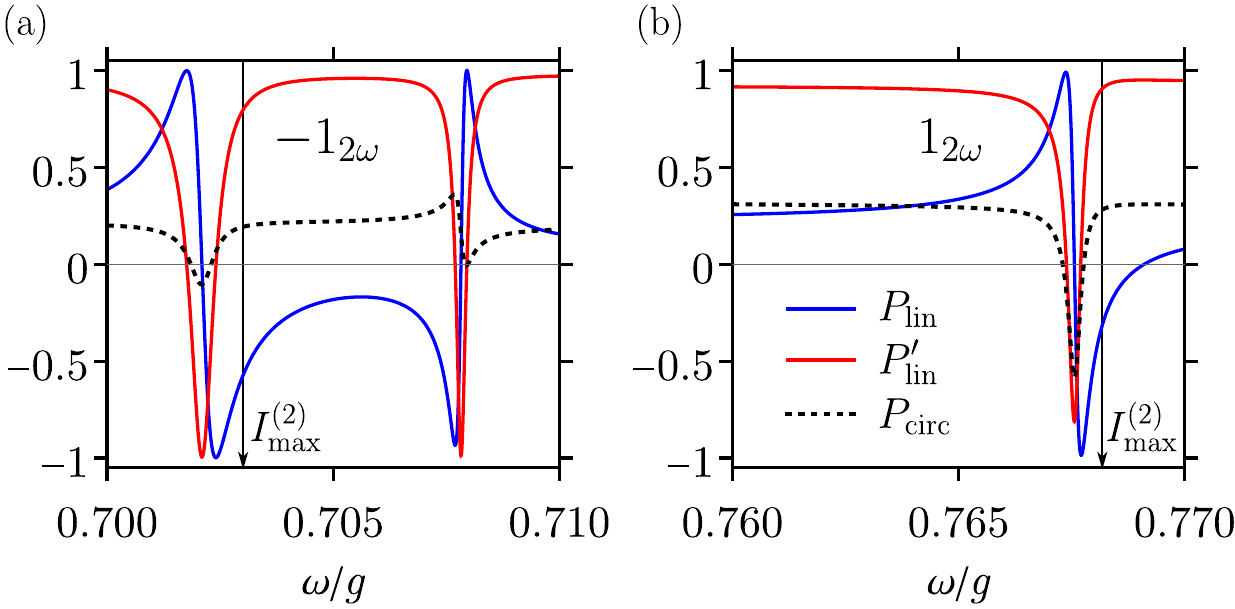}
    \caption{The Stokes polarization parameters $P_{\rm lin}$, $P'_{\rm lin}$, and $P_{\rm circ}$ of the diffracted beams ''$-1_{2\omega}$'' and ''$+1_{2\omega}$'' as a function of the incident light frequency in the vicinity of resonances. The vertical arrows show the
    spectral positions of the beam intensity maxima. The curves are calculated for the same parameters as Fig.~\ref{fig:SHG-Difr-Int} and $\chi_{yyy}=\chi_{xyy}$.}
    \label{fig:SHG-Difr-Pol}
\end{figure}

In conclusion of the discussion of the diffracted beams, we note that their polarization  depends on the parameters of the nonlinear susceptibility and can vary across the spectrum. As an example, Figure~\ref{fig:SHG-Difr-Pol} shows the polarization of the beams ''$-1_{2\omega}$'' and ''$+1_{2\omega}$'', expressed in terms of the Stokes parameters $P_{\rm lin}$, $P'_{\rm lin}$, and $P_{\rm circ}$, as a function of the incident light frequency in the vicinity of resonances. The curves are calculated for $\chi_{yyy}=\chi_{xyy}$. Far from the resonances, the beams are mostly linearly polarized with $P'_{\rm lin} \approx 1$, which follows from $\chi_{yyy}=\chi_{xyy}$. An admixture of the circular polarization $P_{\rm circ}$ is present due to the oblique propagation of the diffraction beams. At the resonances, the amplitudes and the phases of the Fourier harmonics of the near field vary, which in turn modifies the polarization of the emitted radiation. When passing through the resonance, the linear polarization rotates by $180^\circ$.

\section{summary}

We have developed an analytical theory of second harmonic generation in the hybrid structure
consisting of a nonlinear 2D material integrated with a metasurface waveguide.
Using the scattering formalism, we have calculated the amplification spectra and polarizations of the forward emitted and diffracted second harmonic beams at the meta-waveguide resonances associated with both leaky and quasi-BIC modes. The forward emitted beam originates from the second-order nonlinear susceptibility of the 2D material while the diffracted beams contain also a contribution of the nonlocal nonlinear response originating from the spatial inhomogeneity of the electromagnetic field in the 2D plane. The nonlocal mechanism gives rise to the second harmonic beams even if both the 2D material and meta-waveguide are centrosymmetric. 

The developed analytical approach captures the resonant contributions to the second harmonic emission, accurately describing the amplitudes, widths, and spectral positions of the resonances in terms of the Fourier harmonics of the meta-waveguide polarizability $\alpha_n$.
For low-contrast gratings ($|\alpha_n| \ll |\alpha_0|$ for $n \neq 0$), the full harmonic description of the near electromagnetic field can be transformed into the model of two coupled near field harmonics, where the interaction with other harmonics leads to renormalization of the parameters. This model describes quite well both the leaky and quasi-BIC resonances as well as their interaction and interconversion.
In low-loss structures, the factor of enhancement of the intensity of second harmonic emission
at leaky resonances is determined by radiative broadening. For the first leaky resonance at the normal incidence of light, the factor can be estimated as $\sim (\Re \alpha_0/\alpha_1)^4$, where $\alpha_0$ is the mean polarizability and $\alpha_1$ is the first Fourier harmonic. 
Symmetry breaking due to deviation of the light incidence from normal or 
asymmetry in the meta-waveguide enables the excitation of the complementary quasi-BIC mode. 
The spectral shift of the quasi-BIC resonance from the leaky resonance is determined mainly by the second harmonic of the polarizability $\alpha_2$ and can be tuned to a positive or negative value by adjusting the meta-waveguide parameters.
The radiative broadening of the quasi-BIC mode can be vanishingly small so that the
enhancement of second harmonic generation is ultimately limited by inhomogeneous broadening and absorption in the system. The maximal enhancement factor at the quasi-BIC resonance reaches 
$\sim(\Re \alpha_0/ \Im \alpha_0)^4$ at the angle $\theta \sim \sqrt{\Im \alpha_0/ \Re \alpha_0}$ of the light incidence.

\acknowledgments

This work was supported by the Russian Science Foundation (Project No. 22-12-00211-$\Pi$). E.S.V. acknowledges also the support by the Foundation for the Advancement of Theoretical Physics and Mathematics "BASIS".

\bibliography{SHG}

@article{Bulgakov2018,
  title = {Avoided crossings and bound states in the continuum in low-contrast dielectric gratings},
  author = {Bulgakov, Evgeny N. and Maksimov, Dmitrii N.},
  journal = {Phys. Rev. A},
  volume = {98},
  issue = {5},
  pages = {053840},
  numpages = {7},
  year = {2018},
  month = {Nov},
  publisher = {American Physical Society},
  doi = {10.1103/PhysRevA.98.053840},
  url = {https://link.aps.org/doi/10.1103/PhysRevA.98.053840}
}

@article{Ning2023,
url = {https://doi.org/10.1515/nanoph-2023-0124},
title = {Giant enhancement of second harmonic generation from monolayer {2D} materials placed on photonic moir{\'e} superlattice},
author = {Tingyin Ning and Lina Zhao and Yanyan Huo and Yangjian Cai and Yingying Ren},
pages = {4009--4016},
volume = {12},
number = {21},
journal = {Nanophotonics},
doi = {doi:10.1515/nanoph-2023-0124},
year = {2023},
lastchecked = {2025-05-27}
}

@article{Koshelev2020,
  title = {Subwavelength dielectric resonators for nonlinear nanophotonics},
  volume = {367},
  ISSN = {1095-9203},
  url = {http://dx.doi.org/10.1126/science.aaz3985},
  DOI = {10.1126/science.aaz3985},
  number = {6475},
  journal = {Science},
  publisher = {American Association for the Advancement of Science (AAAS)},
  author = {Koshelev,  Kirill and Kruk,  Sergey and Melik-Gaykazyan,  Elizaveta and Choi,  Jae-Hyuck and Bogdanov,  Andrey and Park,  Hong-Gyu and Kivshar,  Yuri},
  year = {2020},
  month = jan,
  pages = {288–292}
}

@article{Gunyaga2025,
  title = {Second Harmonic Generation due to the Spatial Structure of a Radiation Beam},
  author = {Gunyaga, A. A. and Durnev, M. V. and Tarasenko, S. A.},
  journal = {Phys. Rev. Lett.},
  volume = {134},
  issue = {15},
  pages = {156901},
  numpages = {7},
  year = {2025},
  month = {Apr},
  publisher = {American Physical Society},
  doi = {10.1103/PhysRevLett.134.156901},
  url = {https://link.aps.org/doi/10.1103/PhysRevLett.134.156901}
}

@article{Koshelev2021,
  title = {Bound states in the continuum in photonic structures},
  volume = {66},
  ISSN = {1468-4780},
  url = {http://dx.doi.org/10.3367/UFNe.2021.12.039120},
  DOI = {10.3367/ufne.2021.12.039120},
  number = {05},
  journal = {Physics-Uspekhi},
  publisher = {Uspekhi Fizicheskikh Nauk (UFN) Journal},
  author = {Koshelev,  Kirill L. and Sadrieva,  Zarina F. and Shcherbakov,  Alexey A. and Kivshar,  Yu.S. and Bogdanov,  Andrey A.},
  year = {2021},
  month = dec,
  pages = {494–517}
}

@article{Tikhodeev2002,
  title = {Quasiguided modes and optical properties of photonic crystal slabs},
  author = {Tikhodeev, S. G. and Yablonskii, A. L. and Muljarov, E. A. and Gippius, N. A. and Ishihara, Teruya},
  journal = {Phys. Rev. B},
  volume = {66},
  issue = {4},
  pages = {045102},
  numpages = {17},
  year = {2002},
  month = {Jul},
  publisher = {American Physical Society},
  doi = {10.1103/PhysRevB.66.045102},
  url = {https://link.aps.org/doi/10.1103/PhysRevB.66.045102}
}

@article{Vyatkin2025,
  title = {Emergent spin and orbital angular momentum of light in twisted photonic bilayer},
  author = {Vyatkin, Egor S. and Poshakinskiy, Alexander V. and Tarasenko, Sergey A.},
  journal = {Phys. Rev. B},
  volume = {111},
  issue = {12},
  pages = {125303},
  numpages = {9},
  year = {2025},
  month = {Mar},
  publisher = {American Physical Society},
  doi = {10.1103/PhysRevB.111.125303},
  url = {https://link.aps.org/doi/10.1103/PhysRevB.111.125303}
}

@article{Liu2016,
author = {Liu, Sheng and Sinclair, Michael B. and Saravi, Sina and Keeler, Gordon A. and Yang, Yuanmu and Reno, John and Peake, Gregory M. and Setzpfandt, Frank and Staude, Isabelle and Pertsch, Thomas and Brener, Igal},
title = {Resonantly Enhanced Second-Harmonic Generation Using {III–V} Semiconductor All-Dielectric Metasurfaces},
journal = {Nano Letters},
volume = {16},
number = {9},
pages = {5426-5432},
year = {2016},
doi = {10.1021/acs.nanolett.6b01816},
URL = {https://doi.org/10.1021/acs.nanolett.6b01816},

}

@article{Koshelev2018,
  title = {Asymmetric Metasurfaces with High-{Q} Resonances Governed by Bound States in the Continuum},
  author = {Koshelev, Kirill and Lepeshov, Sergey and Liu, Mingkai and Bogdanov, Andrey and Kivshar, Yuri},
  journal = {Phys. Rev. Lett.},
  volume = {121},
  issue = {19},
  pages = {193903},
  numpages = {6},
  year = {2018},
  month = {Nov},
  publisher = {American Physical Society},
  doi = {10.1103/PhysRevLett.121.193903},
  url = {https://link.aps.org/doi/10.1103/PhysRevLett.121.193903}
}

@article{Xu2019,
author = {Xu, Lei and Zangeneh Kamali, Khosro and Huang, Lujun and Rahmani, Mohsen and Smirnov, Alexander and Camacho-Morales, Rocio and Ma, Yixuan and Zhang, Guoquan and Woolley, Matt and Neshev, Dragomir and Miroshnichenko, Andrey E.},
title = {Dynamic Nonlinear Image Tuning through Magnetic Dipole Quasi-{BIC} Ultrathin Resonators},
journal = {Advanced Science},
volume = {6},
number = {15},
pages = {1802119},
doi = {https://doi.org/10.1002/advs.201802119},
url = {https://advanced.onlinelibrary.wiley.com/doi/abs/10.1002/advs.201802119},
year = {2019}
}

@article{Dyakov2021,
author = {Dyakov, Sergey A. and Stepikhova, Margarita V. and Bogdanov, Andrey A. and Novikov, Alexey V. and Yurasov, Dmitry V. and Shaleev, Mikhail V. and Krasilnik, Zakhary F. and Tikhodeev, Sergei G. and Gippius, Nikolay A.},
title = {Photonic Bound States in the Continuum in {S}i Structures with the Self-Assembled {G}e Nanoislands},
journal = {Laser \& Photonics Reviews},
volume = {15},
number = {7},
pages = {2000242},
doi = {https://doi.org/10.1002/lpor.202000242},
url = {https://onlinelibrary.wiley.com/doi/abs/10.1002/lpor.202000242},
year = {2021}
}

@article{Bernhardt2020,
author = {Bernhardt, Nils and Koshelev, Kirill and White, Simon J.U. and Meng, Kelvin Wong Choon and Fr{\"o}ch, Johannes E. and Kim, Sejeong and Tran, Toan Trong and Choi, Duk-Yong and Kivshar, Yuri and Solntsev, Alexander S.},
title = {Quasi-{BIC} Resonant Enhancement of Second-Harmonic Generation in $\text{WS}_2$ Monolayers},
journal = {Nano Letters},
volume = {20},
number = {7},
pages = {5309-5314},
year = {2020},
doi = {10.1021/acs.nanolett.0c01603},
URL = {https://doi.org/10.1021/acs.nanolett.0c01603
},
}

@article{Ling2025,
author = {Ling, Haonan and Tang, Yuankai and Tian, Xinyu and Shafirin, Pavel and Hossain, Mozakkar and Vabishchevich, Polina P. and Harutyunyan, Hayk and Davoyan, Artur R.},
title = {Nonlinear van der {W}aals Metasurfaces with Resonantly Enhanced Light Generation},
journal = {Nano Letters},
volume = {25},
number = {23},
pages = {9229-9236},
year = {2025},
doi = {10.1021/acs.nanolett.5c00952},
URL = {https://doi.org/10.1021/acs.nanolett.5c00952},
}

@article{Dean2009,
	author = {Dean,Jesse J. and van Driel,Henry M.},
	journal = {Applied Physics Letters},
	langid = {english},
	number = {26},
	pages = {261910},
	title = {Second harmonic generation from graphene and graphitic films},
	url = {https://doi.org/10.1063/1.3275740},
	volume = {95},
	year = {2009}}

@article{Li2013,
	author = {Li, Yilei and Rao, Yi and Mak, Kin Fai and You, Yumeng and Wang, Shuyuan and Dean, Cory R. and Heinz, Tony F.},
	date = {2013/07/10},
	doi = {10.1021/nl401561r},
	isbn = {1530-6984},
	journal = {Nano Letters},
	journal1 = {Nano Letters},
	journal2 = {Nano Lett.},
	langid = {english},
	number = {7},
	pages = {3329--3333},
	publisher = {American Chemical Society},
	title = {Probing Symmetry Properties of Few-Layer $\text{MoS}_2$ and {h-BN} by Optical Second-Harmonic Generation},
	type = {doi: 10.1021/nl401561r},
	url = {https://doi.org/10.1021/nl401561r},
	volume = {13},
	year = {2013},
	year1 = {2013}}

@article{Mennel2018,
    author = {Mennel, Lukas and Paur, Matthias and Mueller, Thomas},
    title = {Second harmonic generation in strained transition metal dichalcogenide monolayers: $\text{MoS}_2$, $\text{MoSe}_2$, $\text{WS}_2$, and $\text{WSe}_2$},
    journal = {APL Photonics},
    volume = {4},
    number = {3},
    pages = {034404},
    year = {2018},
    month = {12},
    issn = {2378-0967},
    doi = {10.1063/1.5051965},
    url = {https://doi.org/10.1063/1.5051965},
}

@article{Zhou2020,
	article-number = {2263},
	author = {Zhou, Linlin and Fu, Huange and Lv, Ting and Wang, Chengbo and Gao, Hui and Li, Daqian and Deng, Leimin and Xiong, Wei},
	doi = {10.3390/nano10112263},
	issn = {2079-4991},
	journal = {Nanomaterials},
	langid = {english},
	number = {11},
	pages = {2263},
	pubmedid = {33207552},
	title = {Nonlinear Optical Characterization of {2D} Materials},
	url = {https://www.mdpi.com/2079-4991/10/11/2263},
	volume = {10},
	year = {2020}}

@article{Stepanov2020,
	author = {Stepanov, E. A. and Semin, S. V. and Woods, C. R. and Vandelli, M. and Kimel, A. V. and Novoselov, K. S. and Katsnelson, M. I.},
	date = {2020/06/17},
	doi = {10.1021/acsami.0c05965},
	isbn = {1944-8244},
	journal = {ACS Applied Materials \& Interfaces},
	journal1 = {ACS Applied Materials \& Interfaces},
	journal2 = {ACS Appl. Mater. Interfaces},
	langid = {english},
	month = {06},
	number = {24},
	pages = {27758--27764},
	publisher = {American Chemical Society},
	title = {Direct Observation of Incommensurate--Commensurate Transition in Graphene-{hBN} Heterostructures via Optical Second Harmonic Generation},
	type = {doi: 10.1021/acsami.0c05965},
	url = {https://doi.org/10.1021/acsami.0c05965},
	volume = {12},
	year = {2020},
	year1 = {2020}}

@article{Wang2009,
	author = {Wang, Fu Xiang and Rodriguez, Francisco J. and Albers, Willem M. and Ahorinta, Risto and Sipe, J. E. and Kauranen, Martti},
	doi = {10.1103/PhysRevB.80.233402},
	issue = {23},
	journal = {Phys. Rev. B},
	langid = {english},
	numpages = {4},
	pages = {233402},
	publisher = {American Physical Society},
	title = {Surface and bulk contributions to the second-order nonlinear optical response of a gold film},
	url = {https://link.aps.org/doi/10.1103/PhysRevB.80.233402},
	volume = {80},
	year = {2009}}

@article{Golub2014,
  title = {Valley polarization induced second harmonic generation in graphene},
  author = {Golub, L. E. and Tarasenko, S. A.},
  journal = {Phys. Rev. B},
  volume = {90},
  issue = {20},
  pages = {201402},
  numpages = {4},
  year = {2014},
  month = {Nov},
  publisher = {American Physical Society},
  doi = {10.1103/PhysRevB.90.201402},
  url = {https://link.aps.org/doi/10.1103/PhysRevB.90.201402}
}

@article{Durnev2022,
	author = {Durnev, M. V. and Tarasenko, S. A.},
	doi = {10.1103/PhysRevB.106.125426},
	issue = {12},
	journal = {Phys. Rev. B},
	month = {Sep},
	numpages = {9},
	pages = {125426},
	publisher = {American Physical Society},
	title = {Second harmonic generation at the edge of a two-dimensional electron gas},
	url = {https://link.aps.org/doi/10.1103/PhysRevB.106.125426},
	volume = {106},
	year = {2022}}

@article{Hsu2014,
	author = {Hsu, Wei-Ting and Zhao, Zi-Ang and Li, Lain-Jong and Chen, Chang-Hsiao and Chiu, Ming-Hui and Chang, Pi-Shan and Chou, Yi-Chia and Chang, Wen-Hao},
	date = {2014/03/25},
	doi = {10.1021/nn500228r},
	isbn = {1936-0851},
	journal = {ACS Nano},
	journal1 = {ACS Nano},
	journal2 = {ACS Nano},
	month = {03},
	number = {3},
	pages = {2951--2958},
	publisher = {American Chemical Society},
	title = {Second Harmonic Generation from Artificially Stacked Transition Metal Dichalcogenide Twisted Bilayers},
	type = {doi: 10.1021/nn500228r},
	url = {https://doi.org/10.1021/nn500228r},
	volume = {8},
	year = {2014},
	year1 = {2014}}

@article{Yang2020,
	author = {Fuyi Yang and Wenshen Song and Fanhao Meng and Fuchuan Luo and Shuai Lou and Shuren Lin and Zilun Gong and Jinhua Cao and Edward S. Barnard and Emory Chan and Li Yang and Jie Yao},
	doi = {https://doi.org/10.1016/j.matt.2020.08.018},
	issn = {2590-2385},
	journal = {Matter},
	number = {4},
	pages = {1361-1376},
	title = {Tunable Second Harmonic Generation in Twisted Bilayer Graphene},
	url = {https://www.sciencedirect.com/science/article/pii/S2590238520304458},
	volume = {3},
	year = {2020}}

@article{Yao2021,
	author = {Kaiyuan Yao and Nathan R. Finney and Jin Zhang and Samuel L. Moore and Lede Xian and Nicolas Tancogne-Dejean and Fang Liu and Jenny Ardelean and Xinyi Xu and Dorri Halbertal and K. Watanabe and T. Taniguchi and Hector Ochoa and Ana Asenjo-Garcia and Xiaoyang Zhu and D. N. Basov and Angel Rubio and Cory R. Dean and James Hone and P. James Schuck},
	doi = {10.1126/sciadv.abe8691},
	journal = {Science Advances},
	number = {10},
	pages = {eabe8691},
	title = {Enhanced tunable second harmonic generation from twistable interfaces and vertical superlattices in boron nitride homostructures},
	url = {https://www.science.org/doi/abs/10.1126/sciadv.abe8691},
	volume = {7},
	year = {2021}}

@article{Paradisanos2022,
	author = {Paradisanos, Ioannis and Raven, Andres Manuel Saiz and Amand, Thierry and Robert, Cedric and Renucci, Pierre and Watanabe, Kenji and Taniguchi, Takashi and Gerber, Iann C. and Marie, Xavier and Urbaszek, Bernhard},
	doi = {10.1103/PhysRevB.105.115420},
	issue = {11},
	journal = {Phys. Rev. B},
	month = {Mar},
	numpages = {9},
	pages = {115420},
	publisher = {American Physical Society},
	title = {Second harmonic generation control in twisted bilayers of transition metal dichalcogenides},
	url = {https://link.aps.org/doi/10.1103/PhysRevB.105.115420},
	volume = {105},
	year = {2022}}

@article{Glazov2011,
	author = {Glazov, M. M.},
	date = {2011/06/01},
	doi = {10.1134/S0021364011070046},
	id = {Glazov2011},
	isbn = {1090-6487},
	journal = {JETP Letters},
	langid = {english},
	number = {7},
	pages = {366--371},
	title = {Second harmonic generation in graphene},
	url = {https://doi.org/10.1134/S0021364011070046},
	volume = {93},
	year = {2011}
	}

@article{Hendry2010,
	author = {Hendry, E. and Hale, P. J. and Moger, J. and Savchenko, A. K. and Mikhailov, S. A.},
	doi = {10.1103/PhysRevLett.105.097401},
	issue = {9},
	journal = {Phys. Rev. Lett.},
	langid = {english},
	month = {Aug},
	numpages = {4},
	pages = {097401},
	publisher = {American Physical Society},
	title = {Coherent Nonlinear Optical Response of Graphene},
	url = {https://link.aps.org/doi/10.1103/PhysRevLett.105.097401},
	volume = {105},
	year = {2010}}

@article{Kuznetsov2016,
author = {Arseniy I. Kuznetsov  and Andrey E. Miroshnichenko  and Mark L. Brongersma  and Yuri S. Kivshar  and Boris Luk’yanchuk },
title = {Optically resonant dielectric nanostructures},
journal = {Science},
volume = {354},
number = {6314},
pages = {aag2472},
year = {2016},
doi = {10.1126/science.aag2472},
URL = {https://www.science.org/doi/abs/10.1126/science.aag2472}}

@article{Kazarinov1976,
    author = {R. F. Kazarinov and Z. N. Sokolova and R. A. Suris} ,
    title = {Planar distributed-feedback optical resonators} ,
    journal = {Sov. Phys. Tech. Phys.},
    year = {1976},
    volume = {21},
    number = {2},
    pages = {130}
}

@article{Celebrano2015,
  title     = {Mode matching in multiresonant plasmonic nanoantennas for
               enhanced second harmonic generation},
  author    = {Celebrano, Michele and Wu, Xiaofei and Baselli, Milena and
               Gro{\ss}mann, Swen and Biagioni, Paolo and Locatelli, Andrea and
               De Angelis, Costantino and Cerullo, Giulio and Osellame, Roberto
               and Hecht, Bert and Du{\`o}, Lamberto and Ciccacci, Franco and
               Finazzi, Marco},
  journal   = {Nature Nanotech.},
  publisher = {Springer Science and Business Media LLC},
  volume    =  {10},
  number    =  {5},
  pages     = {412-417},
  month     =  {may},
  year      =  {2015},
    doi = {10.1038/nnano.2015.69},
    url = {https://doi.org/10.1038/nnano.2015.69}
}

@article{Huang2023,
title = {Resonant leaky modes in all-dielectric metasystems: Fundamentals and applications},
journal = {Physics Reports},
volume = {1008},
pages = {1-66},
year = {2023},
issn = {0370-1573},
doi = {https://doi.org/10.1016/j.physrep.2023.01.001},
url = {https://www.sciencedirect.com/science/article/pii/S0370157323000030},
author = {Lujun Huang and Lei Xu and David A. Powell and Willie J. Padilla and Andrey E. Miroshnichenko},
}

@article{Grinblat2021,
author = {Grinblat, Gustavo},
title = {Nonlinear Dielectric Nanoantennas and Metasurfaces: Frequency Conversion and Wavefront Control},
journal = {ACS Photonics},
volume = {8},
number = {12},
pages = {3406-3432},
year = {2021},
doi = {10.1021/acsphotonics.1c01356},

URL = { 
    
        https://doi.org/10.1021/acsphotonics.1c01356
    
    

},
}

@article{Wang2025,
  title = {All-dielectric nonlinear metasurface: from visible to vacuum ultraviolet},
  volume = {2},
  ISSN = {2948-216X},
  URL = {https://doi.org/10.1038/s44310-024-00050-5},
  DOI = {10.1038/s44310-024-00050-5},
  number = {1},
  pages = {4},
  journal = {npj Nanophotonics},
  publisher = {Springer Science and Business Media LLC},
  author = {Wang,  Zhihui and Lin,  Rong and Yao,  Jin and Tsai,  Din Ping},
  year = {2025},
  month = {jan} }

@article{Meng2021,
author={Meng, Yuan
and Chen, Yizhen
and Lu, Longhui
and Ding, Yimin
and Cusano, Andrea
and Fan, Jonathan A.
and Hu, Qiaomu
and Wang, Kaiyuan
and Xie, Zhenwei
and Liu, Zhoutian
and Yang, Yuanmu
and Liu, Qiang
and Gong, Mali
and Xiao, Qirong
and Sun, Shulin
and Zhang, Minming
and Yuan, Xiaocong
and Ni, Xingjie},
title={Optical meta-waveguides for integrated photonics and beyond},
journal={Light: Science {\&} Applications},
year={2021},
month={Nov},
day={22},
volume={10},
number={1},
pages={235},
issn={2047-7538},
doi={10.1038/s41377-021-00655-x},
url={https://doi.org/10.1038/s41377-021-00655-x}
}

@article{WangJ2025,
author={Wang, Jun
and Wang, Wei
and Ding, Weiqiang
and Lin, Jie
and Jin, Peng
and Liu, Shutian
and Zhou, Keya},
title={All-dielectric meta-waveguides for on-chip integration},
journal={npj Nanophotonics},
year={2025},
month={May},
day={27},
volume={2},
number={1},
pages={20},
issn={2948-216X},
doi={10.1038/s44310-025-00069-2},
url={https://doi.org/10.1038/s44310-025-00069-2}
}

@article{Salakhova2023,
  title = {Twist-tunable moir\'e optical resonances},
  author = {Salakhova, Natalia S. and Fradkin, Ilia M. and Dyakov, Sergey A. and Gippius, Nikolay A.},
  journal = {Phys. Rev. B},
  volume = {107},
  issue = {15},
  pages = {155402},
  numpages = {9},
  year = {2023},
  month = {Apr},
  publisher = {American Physical Society},
  doi = {10.1103/PhysRevB.107.155402},
  url = {https://link.aps.org/doi/10.1103/PhysRevB.107.155402}
}

@Article{Lou2021,
  title = {Theory for Twisted Bilayer Photonic Crystal Slabs},
  author = {Lou, Beicheng and Zhao, Nathan and Minkov, Momchil and Guo, Cheng and Orenstein, Meir and Fan, Shanhui},
  journal = {Phys. Rev. Lett.},
  volume = {126},
  issue = {13},
  pages = {136101},
  numpages = {6},
  year = {2021},
  month = {Mar},
  publisher = {American Physical Society},
  doi = {10.1103/PhysRevLett.126.136101},
  url = {https://link.aps.org/doi/10.1103/PhysRevLett.126.136101}
}

@article{Zang2021,
title = {Metasurfaces for manipulating terahertz waves},
journal = {Light: Advanced Manufacturing},
volume = {2},
number = {LAM2020080023},
pages = {148},
year = {2021},
issn = {2689-9620},
doi = {10.37188/lam.2021.010},
url = {https://www.light-am.com//article/id/e38c4162-ee77-40b6-9868-3f40b301199e},
author = {Zang, Xiaofei and Yao, Bingshuang and Chen, Lin and Xie, Jingya and Guo, Xuguang and Balakin, Alexei V. and Shkurinov, Alexander P. and Zhuang, Songlin}
}

\end{document}